\documentclass[twocolumn,aps, prb, showpacs,preprintnumbers,amsmath,amssymb]{revtex4}

\usepackage{graphicx}
\usepackage{dcolumn}
\usepackage{bm}
\usepackage[percent]{overpic}

\begin{document}
\newcommand{\Arg}[1]{\mbox{Arg}\left[#1\right]}
\newcommand{\bb}{\mathbf}
\newcommand{\braopket}[3]{\left \langle #1\right| \hat #2 \left|#3 \right \rangle}
\newcommand{\braket}[2]{\langle #1|#2\rangle}
\newcommand{\be}{\[}
\newcommand{\br}{\vspace{4mm}}
\newcommand{\bra}[1]{\langle #1|}
\newcommand{\braketbraket}[4]{\langle #1|#2\rangle\langle #3|#4\rangle}
\newcommand{\braop}[2]{\langle #1| \hat #2}
\newcommand{\dd}[1]{ \! \! \!  \mbox{d}#1\ }
\newcommand{\DD}[2]{\frac{\! \! \! \mbox d}{\mbox d #1}#2}
\renewcommand{\det}[1]{\mbox{det}\left(#1\right)}
\newcommand{\ee}{\]} 
\newcommand{\eg}{\textbf{\\  Example: \ \ \ }}
\newcommand{\Imag}[1]{\mbox{Im}\left(#1\right)}
\newcommand{\ket}[1]{|#1\rangle}
\newcommand{\ketbra}[2]{|#1\rangle \langle #2|}
\newcommand{\kp}{\arccos(\frac{\omega - \epsilon}{2t})}
\newcommand{\ldos}{\mbox{L.D.O.S.}}
\renewcommand{\log}[1]{\mbox{log}\left(#1\right)}
\newcommand{\Log}{\mbox{log}}
\newcommand{\Modsq}[1]{\left| #1\right|^2}
\newcommand{\nb}{\textbf{Note: \ \ \ }}
\newcommand{\op}[1]{\hat {#1}}
\newcommand{\opket}[2]{\hat #1 | #2 \rangle}
\newcommand{\occ}{\mbox{Occ. Num.}}
\newcommand{\Real}[1]{\mbox{Re}\left(#1\right)}
\newcommand{\so}{\Rightarrow}
\newcommand{\sol}{\textbf{Solution: \ \ \ }}
\newcommand{\thetafn}[1]{\  \! \theta \left(#1\right)}
\newcommand{\tin}{\int_{-\infty}^{+\infty}\! \! \!\!\!\!\!}
\newcommand{\Tr}[1]{\mbox{Tr}\left(#1\right)}
\newcommand{\kb}{k_B}
\newcommand{\rad}{\mbox{ rad}}
\preprint{APS/123-QED}

\title{Magnetization profile for impurities in graphene nanoribbons}

\author{S. R. Power$^{(a)}$, V. M. de Menezes$^{(b)}$, S. B. Fagan$^{(c)}$ and M. S. Ferreira$^{(a)}$}
\email{ferreirm@tcd.ie}
\affiliation{%
(a) School of Physics, Trinity College Dublin, Dublin 2, Ireland \\
(b) Departamento de F\'{\i}sica, Universidade Federal de Santa Maria, UFSM, 97105-900, RS, Brazil \\
(c) \'Area de Ci\^encias Tecnol\'ogicas, Centro Universit\'ario Franciscano, UNIFRA, 97010-032, Santa Maria - RS, Brazil
}

\date{\today}

\begin{abstract}
The magnetic properties of graphene-related materials and in particular the spin-polarised edge states predicted for pristine graphene nanoribbons (GNRs) with certain edge geometries have received much attention recently due to a range of possible technological applications. However, the magnetic properties of pristine GNRs are not predicted to be particularly robust in the presence of edge disorder. In this work, we examine the magnetic properties of GNRs doped with transition-metal atoms using a combination of mean-field Hubbard and Density Functional Theory techniques. The effect of impurity location on the magnetic moment of such dopants in GNRs is investigated for the two principal GNR edge geometries - armchair and zigzag. Moment profiles are calculated across the width of the ribbon for both substitutional and adsorbed impurities and regular features are observed for zigzag-edged GNRs in particular. Unlike the case of edge-state induced magnetisation, the moments of magnetic impurities embedded in GNRs are found to be particularly stable in the presence of edge disorder. Our results suggest that the magnetic properties of transition-metal doped GNRs are far more robust than those with moments arising intrinsically due to edge geometry.

\end{abstract}

\pacs{}
                 
\maketitle

\section{Introduction}
\begin{figure}

 \includegraphics[width=0.45\textwidth]{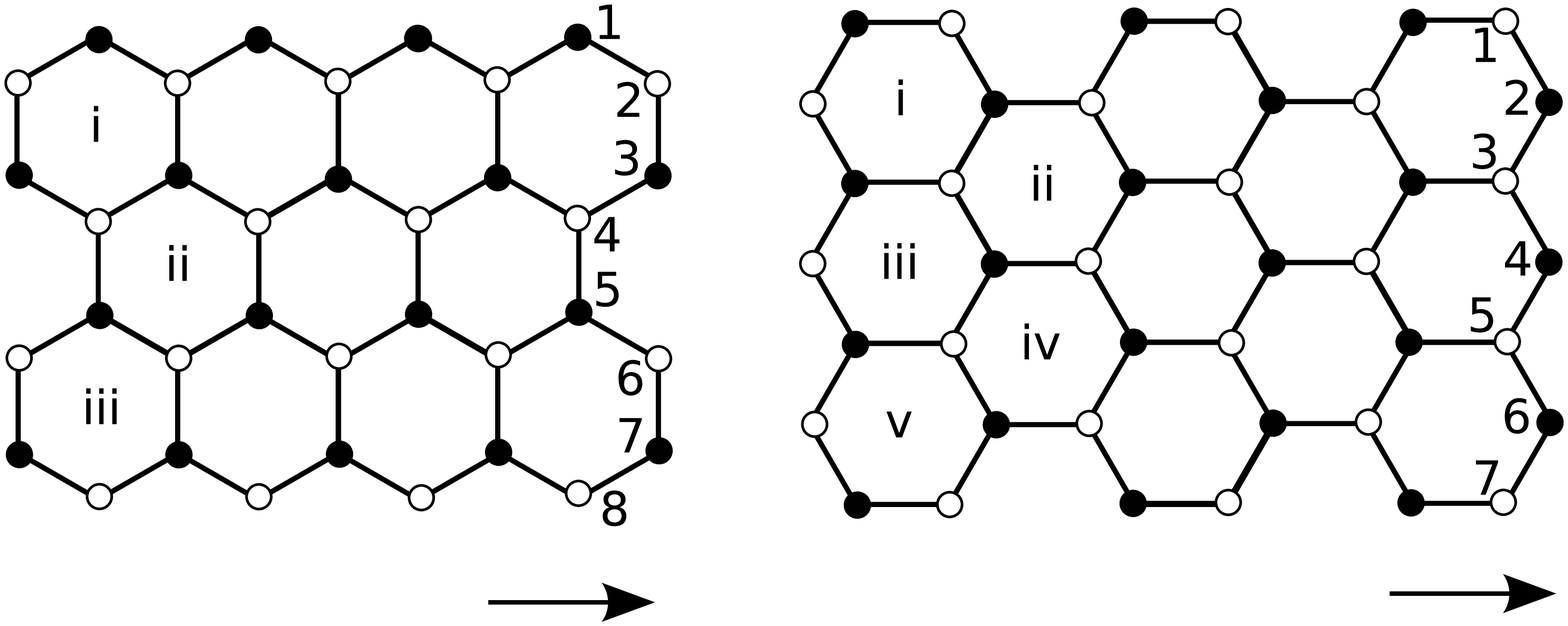}

\vspace{0.8cm}

 \includegraphics[width=0.35\textwidth]{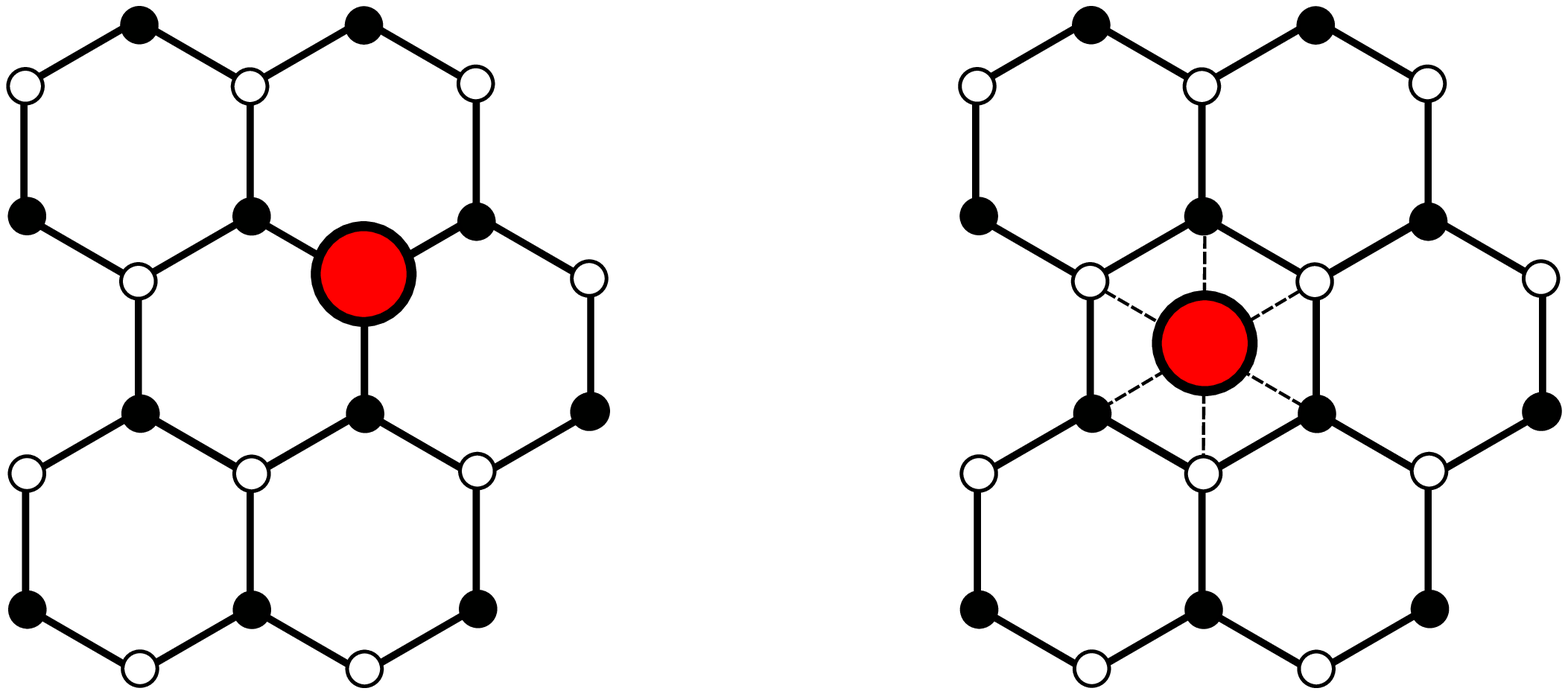}

\caption{Top Panel: Schematic representation of a 4-ZGNR (Zigzag-edged ribbon, left) and 7-AGNR (Armchair-edged ribbon, right). Nanoribbons are labelled by the number, $N$, of zigzag chains (ZGNRs) or dimer lines (AGNRs) across their width. In each case the number of atoms in the repeated unit cell is $2N$. The possible sites for substitutional (centre-adsorbed) impurities are labelled with Arabic (Roman) numerals. The arrows refer to the periodicity direction. \\ Bottom panel: Schematic of a substitutional (left) and centre adsorbed (right) impurity in graphene.}
\label{schematic}
\end{figure}

The experimental discovery of graphene has precipitated wide-ranging research to fully determine the physical properties of this novel material and to pave the way for its application in future technological devices.\cite{Novoselov:2004graphene, Novoselov:2005graphene,Berger:2006epitaxial, Berger:2004epitaxial, riseofgraphene, neto:graphrmp} Of particular interest is the potential for graphene-based spintronic devices to be realised, and thus much attention has been focused on determining the magnetic properties of graphene.\cite{yazyev:review} The existence of spin-polarised edge states, predicted by many theoretical works,\cite{Fujita:zigzagedgestates, Nakada:1996ribbons, Son:halfmetallic, Rossier:zigzag} when a graphene sheet is cut to have a so-called \emph{zigzag} edge geometry has underpinned a large number of these proposed devices. Narrow stripes of graphene, dubbed Graphene Nanoribbons (GNRs), with parallel zigzag edges, predicted to have opposite spin orientation, in particular are proposed.\cite{Son:halfmetallic, Wimmer:device, Kim:device} The other principal nanoribbon geometry, the armchair case, does not display such interesting spin polarised edges. GNRs with zigzag and armchair edge geometries are shown schematically in the top panels of Fig. \ref{schematic}. Despite theoretical advances in the study of GNRs, experimental validation of their properties has so far been inconclusive, due to the difficulty in patterning the edge geometries required for these effects to be observed. Furthermore, the spin-polarised edge state in zigzag-edged GNRs is predicted to be highly dependent on the edge geometry and not particularly robust under the introduction of edge disorder in the form of vacancy defects or impurity atoms.\cite{Kunstmann:unstable} These factors present major obstacles in the path of utilizing the intrinsic magnetic edge states of graphene in experimentally realisable devices. 
Another possibility that has been proposed is the exploitation of defect-driven magnetic moments that arise in graphene.\cite{yazyev:graphenemagnetism2, Palacios:vacancymag} Magnetic moments have been predicted to form around vacancies and other defects in the graphene lattice and the possibility of engineering a ferromagnetic state in graphene from such moments has been suggested. However, such a claim would seem to be restricted by the implications of the Lieb theorem,\cite{Liebtheorem} which states that any such magnetic moments arise from a disparity between the two sublattices of graphene. Large-scale, randomised disorder would tend to minimise such a disparity and prevent the formation of a ferromagnetic state. A third possibility for incorporating graphene in spintronic devices lies in the doping of graphene systems with magnetic impurity atoms. This approach takes advantage of the indirect exchange coupling, often referred to as the RKKY coupling,\cite{RKKY:RK, RKKY:K, RKKY:Y} between magnetic impurities embedded in a graphene system which is mediated by the conduction electrons of the graphene host. Although this coupling is predicted to decay rapidly in graphene sheets,\cite{saremi:graphenerkky, brey:graphenerkky, black:graphenerkky, sherafati:graphenerkky} the quasi-one-dimensional nature of nanoribbons suggests that a much longer range coupling may persist in these materials for certain impurity configurations, in a similar manner to that found in Carbon Nanotubes.\cite{AntonioDavidIEC,David:IEC, vojislav:acs} 

However, unlike carbon nanotubes, whose periodic boundary conditions ensure the equivalence of lattice sites, the physical properties of graphene nanoribbons are expected to display a strong dependence on the location of introduced impurities.\cite{mucciolo:graphenetransportgaps, me:impseg} Indeed, in a previous paper we have shown that the binding energies of impurity atoms depend strongly on their location across a ribbon.\cite{me:impseg} This leads to impurity distributions heavily weighted towards the edge sites - an observation confirmed elsewhere in the literature.\cite{biel:ribbondoping, rigo:Nidopedribbons, Cruz:subdopedribbons, Yu:nidopedribbons,barone:lidiffusion, brito:audiffusion, Kawazoe:dopedgnrs} In addition to the binding energy, the magnitude of the magnetic moment on an impurity atom should also depend on impurity position. We shall demonstrate here that the principal features of this dependence derive from the underlying electronic structure of the GNR host. In section \ref{model} we show how the dependence can be calculated using a simple tight-binding representation of the graphene electronic structure combined with a self-consistent mean field description of the impurity species. In Section \ref{singleimps}, the features of the dependence will be investigated and the effect of the ribbon geometry and the nature of the impurity considered. The reliability of these results will be demonstrated by comparison to a full ab-initio treatment of Mn inpurities embedded into graphene nanoribbons. The nature of these features will be described comprehensively by considering the two principal GNR geometries, armchair and zigzag edges, for the case of two impurity configurations - substitutional atoms replacing a single carbon atom in the graphene lattice, and centre adsorbed atoms sitting in the centre of a hexagon of carbon atoms. These two configurations are shown schematically in the bottom panels of Fig. \ref{schematic}. In Section \ref{edgedis} the effects of edge disorder on the magnitude of the magnetic moments formed at the various impurity sites across the ribbon are examined. We find that an edge vacancy only has a significant effect on moments formed on neighbouring sites, and that the moment profile quickly returns to that of the pure ribbon when we move a few carbon chains away from the vacancy. This illustrates the relative robustness of magnetic moments introduced into graphene systems through transition metal doping compared with those arising intrinsically, which are unstable in the presence of edge disorder.\cite{Kunstmann:unstable}

\section{Model}
\label{model}
The electronic structure of the system is described using a Hubbard-like Hamiltonian $\hat{H} = H_0 + H_{int}$, where $H_0 = \sum_{ij\mu\nu \sigma} \gamma_{ij}^{\mu\nu} \, {\hat c}_{i\mu\sigma}^\dag \, {\hat c}_{j\nu\sigma}$ represents the electronic kinetic energy plus a spin-independent local potential, and $H_{int}$ is the electron-electron interaction term. The operator $\hat{c}_{i\mu\sigma}^{\dag}$ creates an electron with spin $\sigma$ in atomic orbital $\mu$ on site $i$. The graphene electronic structure is described using a single-orbital nearest-neighbour tight-binding model. The low-energy properties of graphene are known to be well reproduced within this framework. The magnetic impurities are described by a five-fold degenerate $d$ band representing a typical transition-metal magnetic atom. Within $H_0$, we set the onsite energy of the carbon atoms to zero. The carbon-carbon nearest-neighbour hopping is $\gamma_{CC} = -2.7eV$, and we set $t = |\gamma_{CC}|$ as our unit of energy. However a correction to the hopping parameter at the edge of armchair ribbons, $\gamma_{CC}^{E} = 1.12 \gamma_{CC}$, is needed in order to achieve the expected semiconducting behaviour for all ribbon widths.\cite{Son:ribbonenergygaps} The exact value of the hopping parameter between the carbon atoms of the graphene lattice and the magnetic impurity atom, $\gamma_{CM}$, will depend on the impurity atom chosen and can be calculated in a number of ways. While the magnitude of this parameter can amplify the value of the magnetic moment, it is not expected to have a significant qualitative effect on the moment profile across the width of the ribbon. 

We assume that $H_{int}$ is an on-site interaction which takes place between electrons occupying the $d$ orbitals of the transition metal impurities only, and is neglected elsewhere. In this case, the matrix elements of the spin-dependent part of the Hamiltonian reduce to $v_{\mu\nu}^{\sigma}=-\frac{1}{2} \Delta_{\mu} \delta_{\mu \nu} \sigma$, where $\Delta_{\mu}$ represents the local exchange splitting associated with orbital $\mu$, and $\sigma = \pm 1$ for $\uparrow$ and $\downarrow$ spin, respectively. For simplicity we shall assume that the onsite effective exchange integrals $U$ are the same for all $d$ orbitals, whence it follows that $\Delta_{\mu}$ is $\mu$-independent and equal to $\Delta = U m$, where $m$ is the local magnetic moment. The value of $m$, and hence $v_{\mu\nu}$, along with the spin-independent contribution to the onsite energies of the transition metal atoms, $\gamma_{MM}$, is calculated self-consistently within the mean field approximation under constraints imposed by the values of the onsite exchange integral $U$ and the d-band occupation $n_d$.
The d-band occupation is calculated at each stage in the self-consistent procedure by an integral over the density of states, which comes directly from the diagonal element of the real-space Green function associated with the GNR-impurity system. This can be calculated in turn from the recursively calculated Green function for the pristine GNR using the Dyson equation.

The DFT calculations were performed by considering a substitutional or adsorbed Mn impurity located at different positions across the width of zigzag and armchair nanoribbons. These calculations were carried out using the SIESTA code\cite{SIESTA:2002} with the systems placed in a supercell so that the calculations were performed using periodic boundary conditions. Double-zeta plus polarization functions were employed and the exchange-correlation function was adjusted using the generalized gradient approximation according to the  parametrization proposed by Perdew, Burke and Ernzerhof.\cite{PBE} The interactions between the ionic cores and the valence electrons were described with normconserving Troullier-Martins pseudopotentials.\cite{troullier} The structural optimizations were performed with the conjugate gradient approximation\cite{SIESTA:2002} until the residual forces were smaller than 0.05 eV/\AA. Care must be taken when performing DFT calculations within a periodic supercell as moment suppression can occur if neighbouring moments prefer an antiferromagnetic alignment, but are forced to adopt a ferromagnetic alignment by the periodicity of the system.\cite{rapidcomm:emergence} The proper emergence of moments in all our calculations suggests that antiferromagnetic alignment is not an issue in this case, an assertion supported by the energetically favourable ferromagnetic alignment between Mn impurities in carbon nanotubes,\cite{DavidSpinValve} a similar material.

Edge disorder is considered by introducing vacancies at edge sites in the ribbon. Within the simple model, these vacancies are introduced by placing a very large on-site potential ($50t$) on the lattice site containing the vacancy. This is a simplified approach that does not take into account lattice deformation effects that may occur when an atom is removed, but is adequate to describe the electronic effects of removing the relevant orbital from that lattice site. It is worth highlighting that this simplification only occurs in the simple model and that the DFT calculations discussed earlier for systems without edge disorder fully account for relaxation.

\section{Magnetic Impurities} 
\label{singleimps}
We present results for two magnetic impurity configurations - the case of a substitutional atom replacing a carbon atom and also that of a centre-adsorbed impurity sitting in the centre of a hexagon of carbon atoms. For the latter case we assume that the hopping parameters connecting the impurity atom to the lattice are equivalent for each of the neighbouring six carbon atoms. The impurity is moved across the finite width of the ribbon and the self-consistent value for $m$ is calculated at each site. We consider both the zigzag and armchair geometries and a comparison is made for all cases with a full Density Functional Theory calculation. For a simpler qualitative comparison, the quantity of interest is the relative fluctuation of the magnitude of the moment around its value at the centre of the ribbon, $m_c$. This is given by $\frac{\Delta m}{m_c} \equiv \frac{m - m_c}{m_c}$. 

\subsection{Substitutional Atoms}

\begin{figure}
\includegraphics[width=0.45\textwidth]{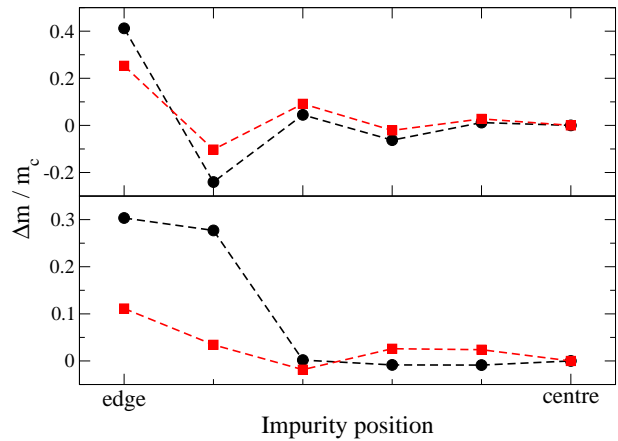}
\caption{The magnetic moment profile across a 6-ZGNR (top panel) and 11-AGNR (bottom panel) for a substitutional impurity, calculated using the self-consistent Hubbard model discussed in the text (red squares)  and also a full DFT treatment with Mn atoms (black circles). An excellent qualitative match is found for the zigzag case, but the armchair match is less convincing. As discussed in the text, this is due to lattice distortion for impurities near the edge of the AGNR. }
\label{figsubs}
\end{figure}

Fig \ref{figsubs} shows the magnetic moment fluctuation as function of impurity location for substitutional impurities across the width of a 6-ZGNR and a 11-ANGR . For the case of zigzag ribbons, we first note an excellent qualitative match between the simple model and the full DFT calculation, from which we infer that the underlying mechanism for the variation in the magnetic moments across the ribbon width is the electronic structure of the nanoribbon. The position dependence arises from quantum interference effects caused by the boundary conditions imposed on the electronic stucture of graphene in the form of the edges of the nanoribbons. Furthermore, we note that the parameters $\gamma_{CM}$, $n_d$ and $U$ which characterise the magnetic species in our simple model can be altered to achieve a better numerical fit, but do not affect the qualitative form of this plot.  The pattern observed is a jagged, sawtooth style curve, characteristic of properties measured across the width of zigzag ribbons and a similar feature can be seen in the binding energies of impurities.\cite{me:impseg} This feature is a sublattice effect which arises from the degeneracy breaking that occurs between the two sublattices of graphene when a zigzag edge is formed. The sublattices are represented schematically in Fig. \ref{schematic} by black or white circles. Each edge of the ribbon is occupied by sites entirely from one of the sublattices, and that sublattice is ``dominant'' on that half of the ribbon. For the case of impurity magnetic moments on zigzag ribbons, this effect manifests itself in creating larger moments on impurities located on the dominant sublattice on either side of the ribbon. In other words, impurity atoms on a black site on the side of the ribbon with black edge sites will have larger moments than their neighbouring white sites. Focusing on a single sublattice, we find that the trend across the ribbon width is for the largest moment to arise on the dominant edge site for that sublattice, to decrease as the impurity is moved towards the centre of the ribbon and to reach its minimum value at the sites neighbouring the opposite edge. We note the all the features discussed here arise in both the simple model and the DFT results, confirming that this is not simply an artifact of our simple model.

The corresponding plots for the armchair case do not agree with each other as convincingly. The tight-binding model is found to underestimate the value of the edge moment found by the DFT calculation. This is because the tight-binding calculation does not take into account the distortions in the honeycomb lattice that arise when a substitutional impurity is introduced near the edge of an AGNR. The relaxed structures (not shown here) for the two impurity sites nearest the ribbon edge are found to be considerably perturbed compared to the pristine ribbon and also to the relaxed structures corresponding to the other impurity sites. The shape of the tight-binding plot for AGNRs is also found to be more dependent on the parameterisation of the impurity than in the zigzag case. This issue will be explored further in the case of centre-adsorbed impurities. This suggests that the moment profile across AGNRs is not as robust as that observed in the ZGNR case, and will vary somewhat according to the magnetic species chosen. However, both tight-binding and DFT models find that the edge impurity sites lead to larger magnetic moments than the central ones.
Fig. \ref{figsubs} also reveals that the sublattice effect noted in zigzag-edged ribbons is absent in the case of armchair edges. This is explained from a cursory inspection of Fig. \ref{schematic} where it is obvious that the degeneracy between black and white lattice sites is unbroken by the imposition of armchair edges. 
The value of the magnetic moment approaches $m_c$ much quicker for AGNRs, and only minor deviations from it are observed away from the edges of the ribbon, whereas in ZGNRS significant deviations are still present deeper into the ribbon.

The dramatic increase observed in the magnetic moment in impurities near the edge of zigzag ribbons is consistent with the presence of a localised edge state at the Fermi energy. This state results in a large peak in the density of states at the Fermi energy. Such a peak provides favourable conditions for moment formation under the Stoner Criterion\cite{rapidcomm:emergence}, and indeed if an intrinsic electron-electron interaction is considered in an undoped ZGNR, will lead to the formation of the spin polarised edges as discussed in the Introduction.

\subsection{Centre adsorbed atoms} 

\begin{figure}
\includegraphics[width=0.45\textwidth]{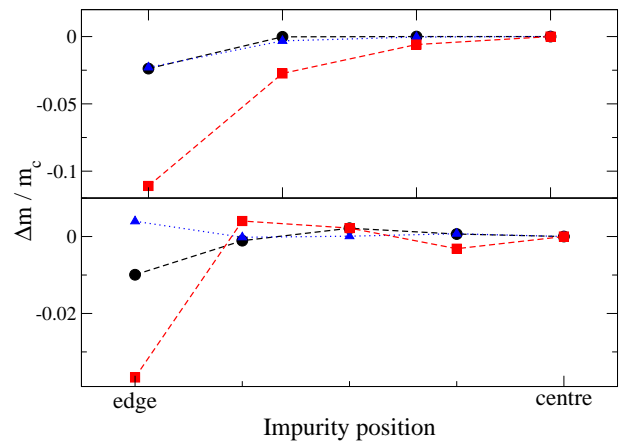}
\caption{The magnetic moment profile across a 8-ZGNR (top panel) and 11-AGNR (bottom panel) for a centre-adsorbed impurity impurity. The black circles show the results from a full DFT treatment with Mn atoms, whereas the red squares and blue triangles show the results from the self-consistent Hubbard model with $\gamma_{CM} = \gamma_{CC}$ and $0.7 \gamma_{CC}$ respectively. We note this parameter does not affect the qualitative features of the moment profile across the ZGNR,  but alters that across the AGNR significantly.}
\label{figcent}
\end{figure}

\begin{figure}
\includegraphics[width=0.23\textwidth, angle=270]{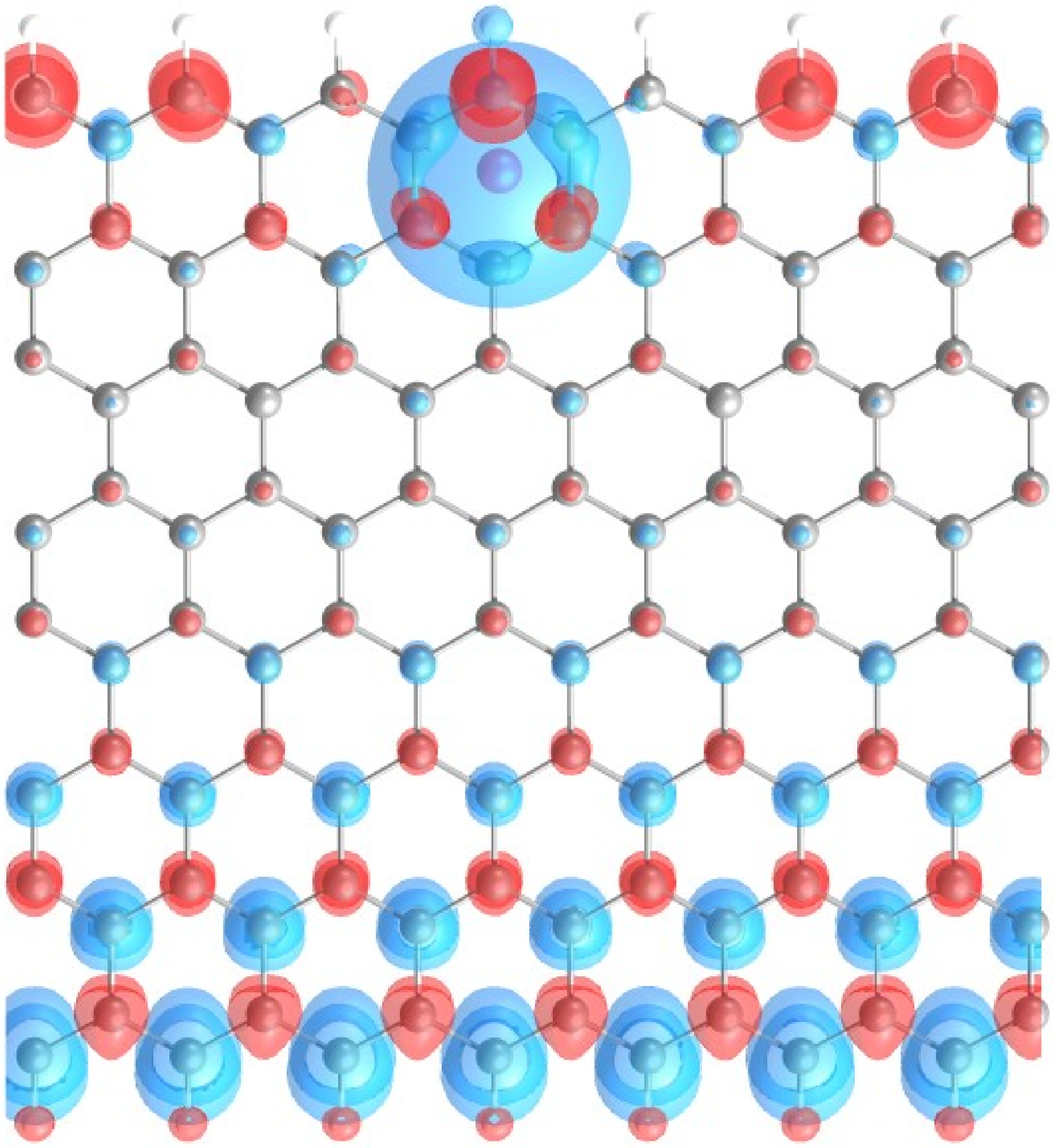} \hspace{0.5cm}
\includegraphics[width=0.23\textwidth, angle=270]{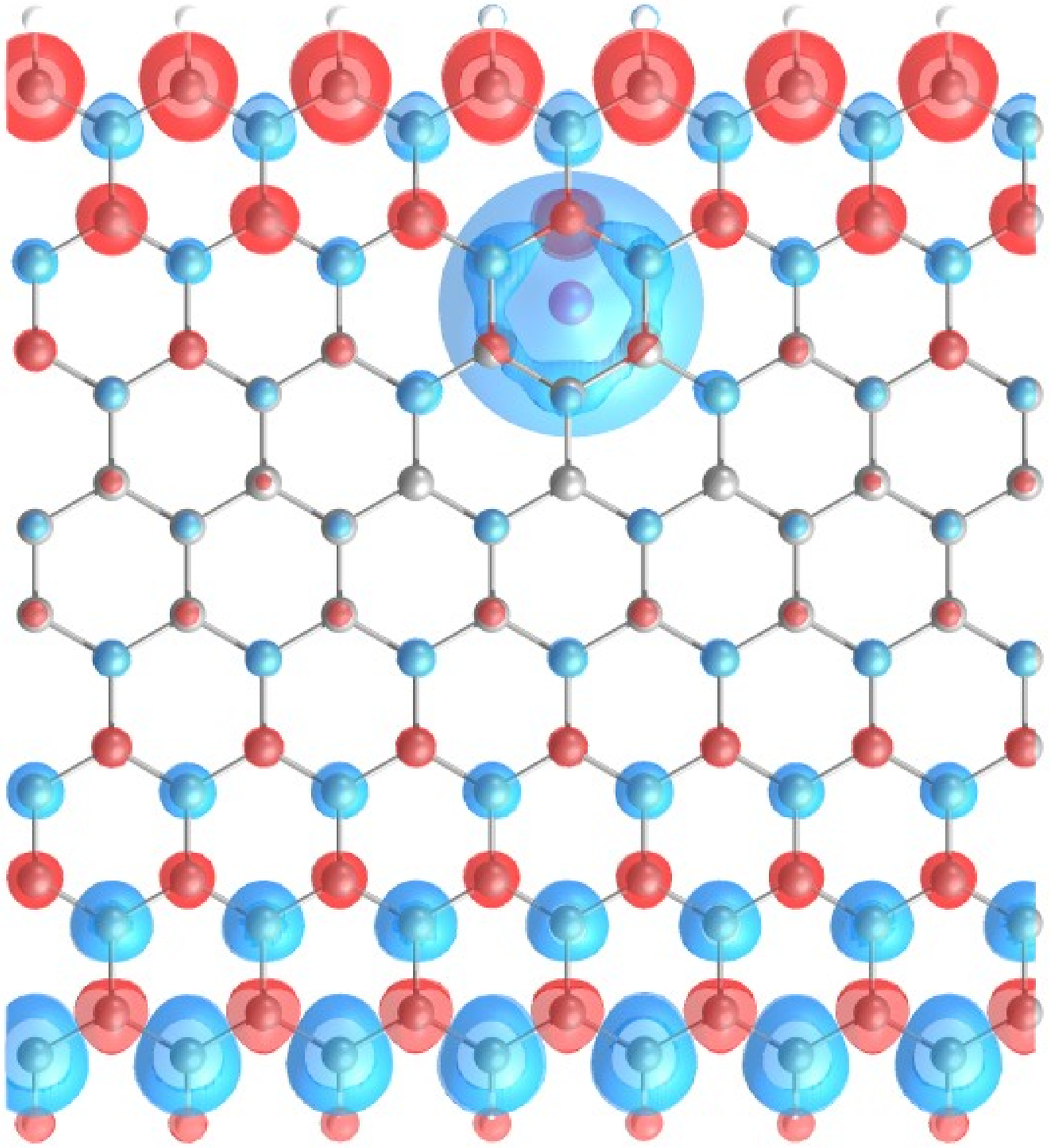}
\caption{Spin density plots showing up (blue) and down (red) spin densities near a centre adsorbed impurity on the edge hexagon (left) and next-to-edge hexagon (right) on a 8-ZGNR. The isosurface used was $0.001 \, e/\mathrm{Bohr}^3$.}
\label{figspinplot}
\end{figure}

Fig. \ref{figcent} shows the magnetic moment fluctuation for a centre-adsorbed impurity at various sites across a GNR, calculated again using both the simple model and a full DFT approach (black cirlces). Within the simple model approach we consider two values of $\gamma_{CM}$, the hopping parameter between the impurity atom and surrounding lattice sites. The values considered are $\gamma_{CM} = \gamma_{CC}$ (red squares) and $\gamma_{CM} = 0.7\gamma_{CC}$ (blue triangles). For the case of ZGNRs (top panel), we note the impressive qualitative match between the models. Furthermore we note that the change in hopping parameter does not effect the qualitative shape of the plot, but can be used to yield a better fit. 
We also note that, unlike the substitutional impurity considered earlier, the sublattice effect is no longer present. This is because the impurity is no longer strongly associated with a particular sublattice, but instead binds to three carbon atoms from each, which has the effect of averaging out any sublattice dependent effects. The general trend of a monotonic increase in the magnetic moment of the Mn impurity is noted as it is moved towards the centre of the ribbon. This is in stark contrast to the result for substitutional impurities, where the largest moment is observed at the edge and, for the dominant sublattice, the moment decreases as the impurity is moved towards the centre of the ribbon. The discrepancy can be explained by the fact that the centre-adsorbed Mn impurity induces fluctuations in the magnetic moments on nearby sites in the graphene lattice. Edge atoms are particularly susceptible to magnetic moments due to the localised state discussed before, and thus have larger induced deviations in their moments than the others. Consequently, centre-adsorbed impurities atoms at the edge of a ZGNR tend to induce large moment deviations on the edge sites, resulting in a smaller moment on the impurity atom itself. This is verified by examining the spin-density plots from the DFT calculation for the case of centre-adsorbed impurities near the edge of a ZGNR. Fig \ref{figspinplot} shows the spin-density plots corresponding to an impurity on the edge hexagon (left) and next-to-edge hexagon (right).
It is clear that the centre-adsorbed impurity nearest the edge introduces a much larger disturbance to the values of the magnetic moments on surrounding sites than the more centrally located impurity. In fact the magnetic edge states are seen to be essentially unperturbed by the latter. In contrast to these DFT calculations, the simple model does not account for the intrinsic magnetic edge states on ZGNRs. However a similar moment profile is recovered as the magnetic impurity atom induces moments, rather than fluctuations of existing moments, on the surrounding lattice sites and these are found to be significantly larger for the case of the centre-adsorbed impurity nearest the ribbon edge.

\begin{figure}
\centering
\begin{tabular}{m{0.19\textwidth} m{0.3\textwidth}}
\centering
\includegraphics[width=0.12\textwidth]{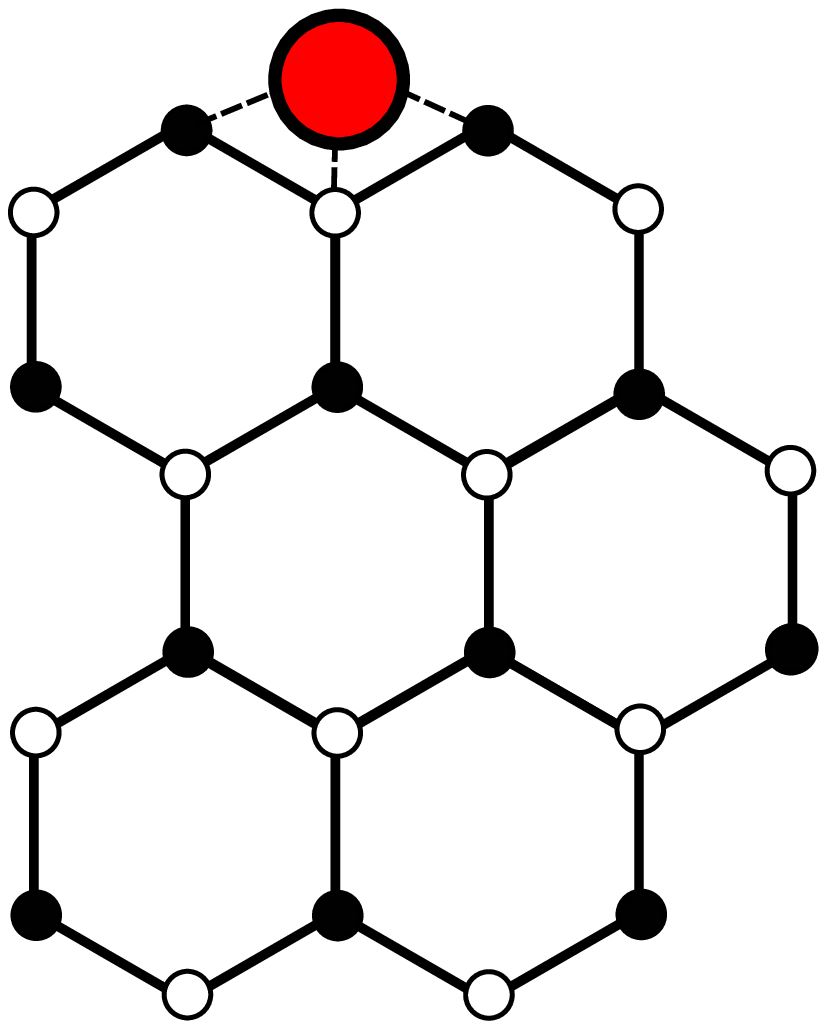}  
\includegraphics[width=0.18\textwidth, angle=270]{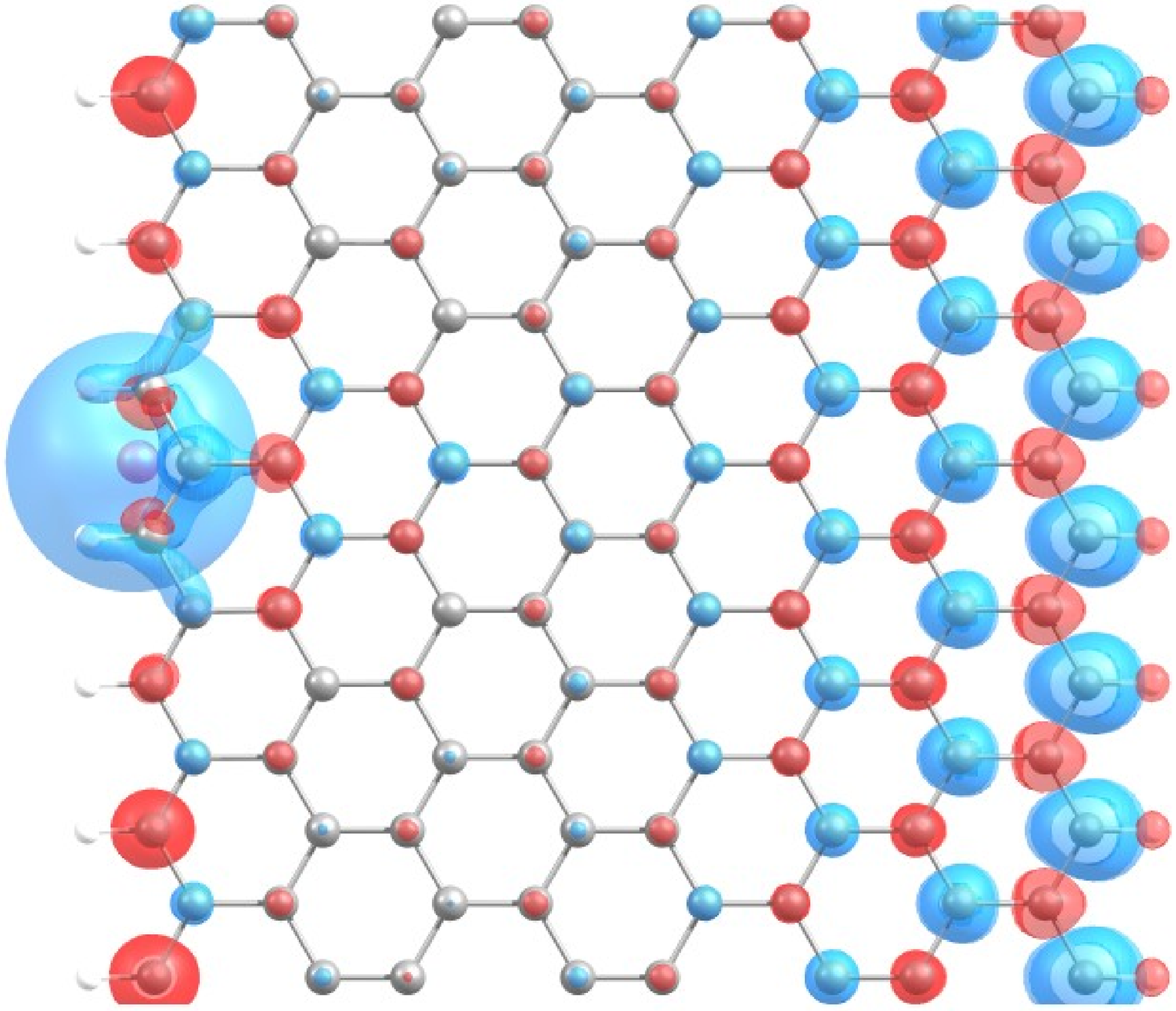}  

&
\vspace{0.5cm} \includegraphics[width=0.27\textwidth]{edgeads} \\
\end{tabular}
 \caption{The edge adsorbed impurity discussed in the text, shown schematically in the top left panel. The spin density plot for this configuration on a 8-ZGNR is shown in the bottom left panel. The right hand side panels show the moment fluctuation (top) and segregation energy function (bottom) for this configuration calculated using the DFT approach, compared to those for the centre-adsorbed locations across the width of a 8-ZGNR.}
\label{zz_edge}
\end{figure}
However, in zigzag ribbons there is an additional type of adsorption site which consists of an impurity atom bound to two edge sites and the site between them. This configuration, which we shall label ``edge-adsorbed'' (EA), is illustrated schematically in Fig. \ref{zz_edge}. It can be viewed as an impurity atom connecting to half a hexagon of the graphene lattice. As this site only occurs at the edge, we cannot study the position dependence of it. However, DFT calculations reveal that a larger moment arises here than for the centre-adsorbed ato locamted nearest the ribbon edge and furthermore that the edge-adsorbed configuration is also more energetically favourable than any of centre-adsorbed sites. This is clear from the right-hand side panels of Fig. \ref{zz_edge}. The upper panel shows the moment fluctuation for the EA case compared with those of the centre adsorbed locations across a 8-ZGNR. The bottom panel plots the segregation energy function, $\beta = \frac{E_B - E_B^c}{|E_B^c|}$, for the same cases. This quantity, introduced in Ref. [\cite{me:impseg}], plots the relative deviation of the binding energy of an impurity on a GNR, $E_B$, around its value at the centre of the ribbon, $ E_B^c$. The edge-adsorbed case is found to be the most energetically favourable. A spin density plot for this type of impurity is shown in the bottom left panel of Fig. \ref{zz_edge} and reveals that this impurity configuration has a less dramatic effect on the moments of nearby edge sites than the centre-adsorbed case on the edge hexagon, consistent with larger moment found for the EA configuration. 

In the results for adsorbed Mn impurities on an AGNR in the bottom panel of Fig. \ref{figcent} we notice that for the DFT result, and the simple model calculation with $\gamma_{CM}=\gamma{CC}$ the edge hexagonal site has a smaller moment on it than the other sites. However the deviation in the value of the edge moment, and indeed of the moment at any site, from $m_c$ is far smaller than in the ZGNR case. The moment profile in this case is essentially flat, with only minor deviations from $m_c$ across the width of the ribbon. Examining the case of $\gamma_{CM}=0.7\gamma_{CC}$ reveals that the shape of the profile across the ribbon is less robust than the zigzag case as the edge moment here is found to be slightly larger than $m_c$. Of the cases examined, the effect is weakest here and does not appear to be very robust. Thus the position dependence of magnetic impurities is smallest for adsorbed impurities on AGNRs and cannot be deemed a significant effect.

\section{Edge Disorder}
\label{edgedis}
In this section the robustness of the features discussed in the previous section will be examined in the presence of edge vacancy defects. This is an important point to consider when comparing impurity-driven magnetic moments in GNRs to those arising intrinsically due the edge states, which have been shown to be particularly vulnerable to edge disorder.\cite{Kunstmann:unstable} For each of the cases discussed in the previous section we examine the effect of a single edge vacancy on the magnitude of a nearby moment, calculated with the mean-field Hubbard model approach. The distance between the magnetic impurity and the edge vacancy is varied to examine the range of this effect. 

\begin{figure}
\hspace{0.1cm}
\includegraphics[width=0.19\textwidth]{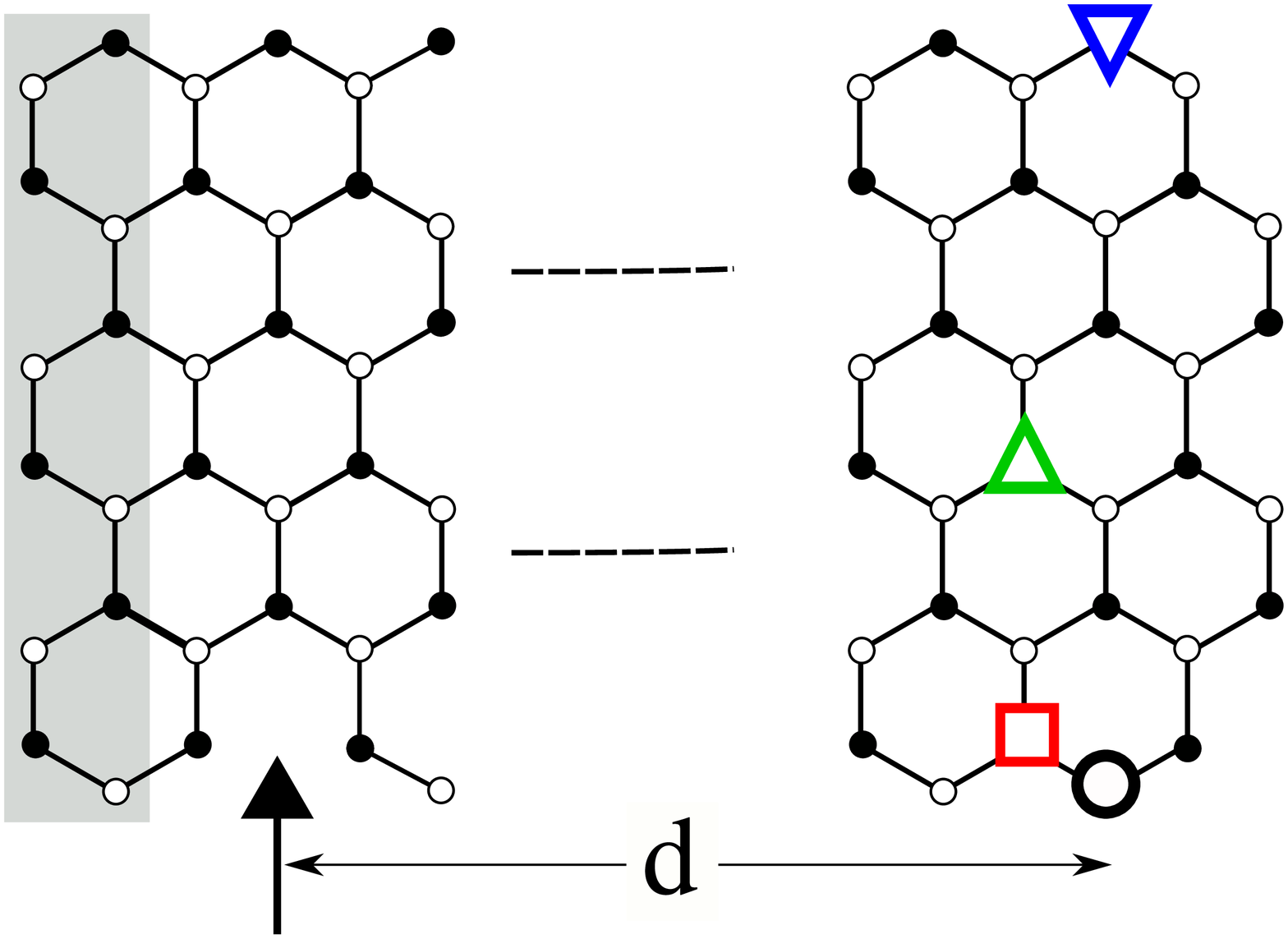}
\hspace{0.2cm}
\includegraphics[width=0.18\textwidth]{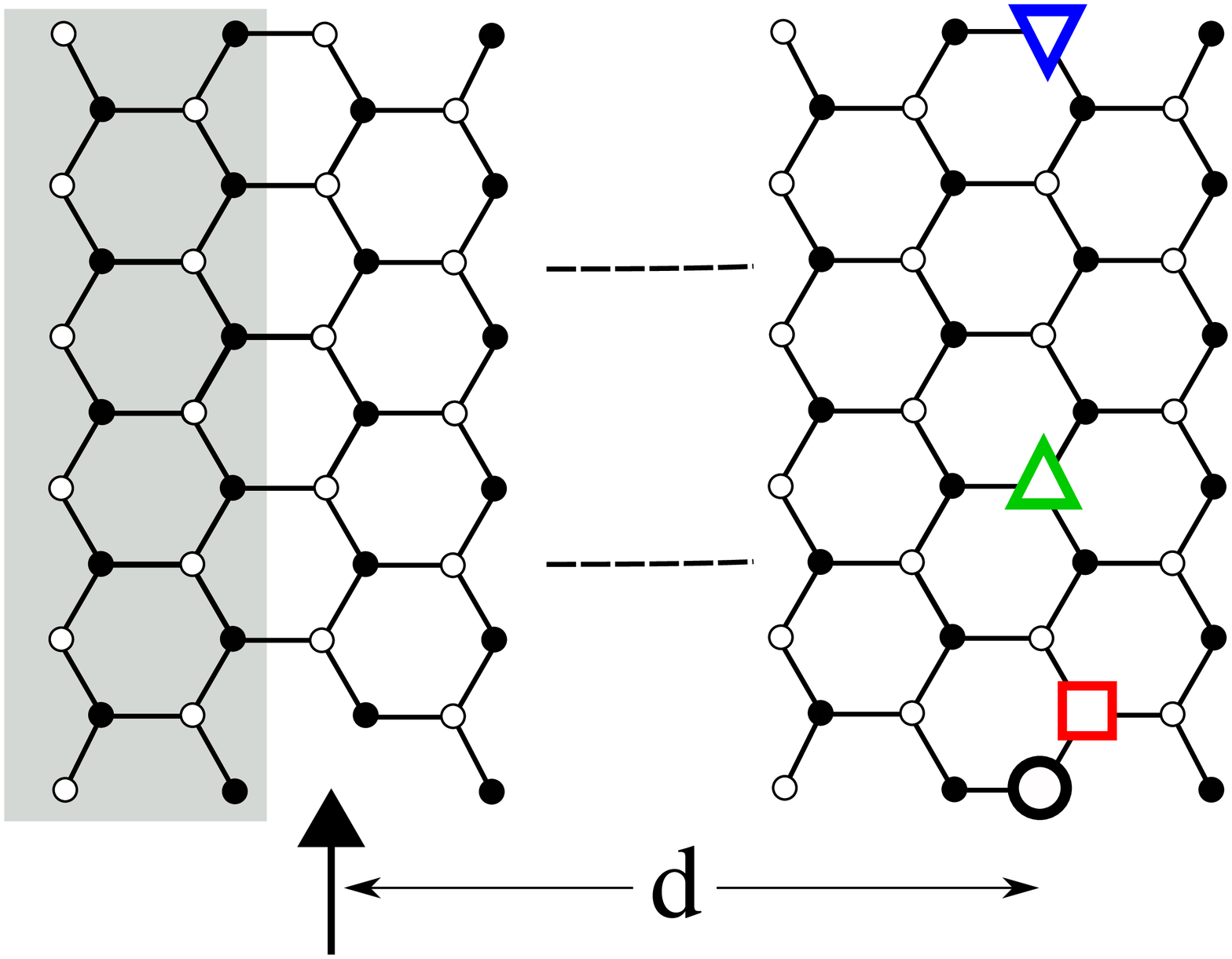}

\vspace{0.2cm}
\includegraphics[width=0.45\textwidth]{edge_fig}

\caption{Effects of a single edge vacancy on the magnetic moments of substitutional magnetic impurities on a 6-ZGNR (left) and 11-AGNR (right). The top images show schematically the edge vacancy and the possible sites of magnetic impurities across the width of the ribbons. For each ribbon we consider sites at the edge (black, circle), next to the edge (red, square), centre (green, triangle) and opposite edge (blue, inverted triangle)  of the ribbon. The graphs underneath show how the fluctuation in the magnetic moment of an impurity at each of these sites under the introduction of an edge vacancy, relative to the moment the impurity would have in the absense of an edge vacancy. This is plotted as a function of distance between the edge vacancy and the unit cell containing the  magnetic impurity, where the distance is given in unit cells. The shaded area in each of the ribbon schematics contains one unit cell of that ribbon.}
\label{edge1}
\end{figure}

\begin{figure}
\begin{tabular}{m{0.25\textwidth} m{0.25\textwidth}} \centering
\hspace{0.8cm}\includegraphics[width=0.16\textwidth]{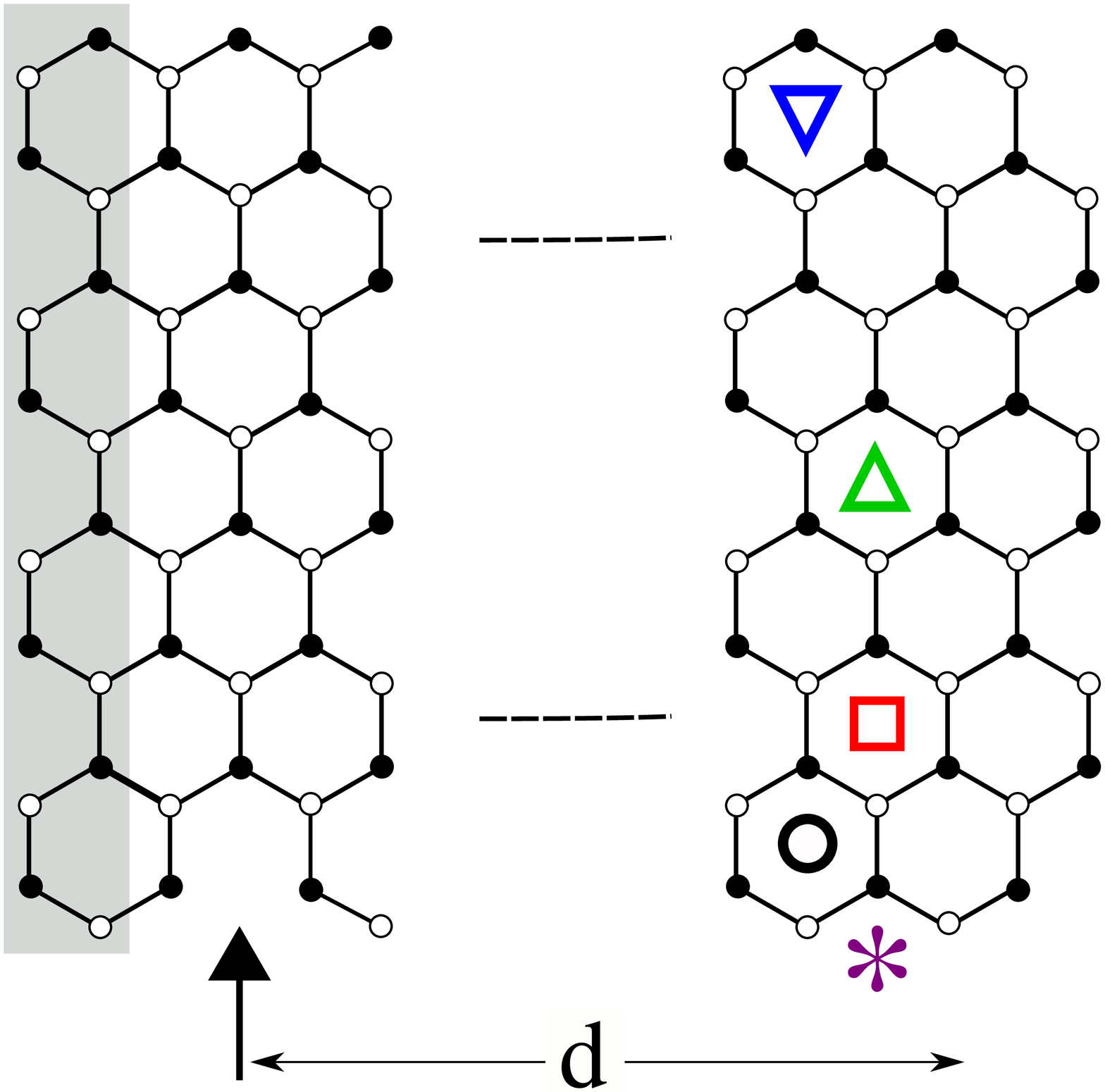}  &
 \includegraphics[width=0.17\textwidth]{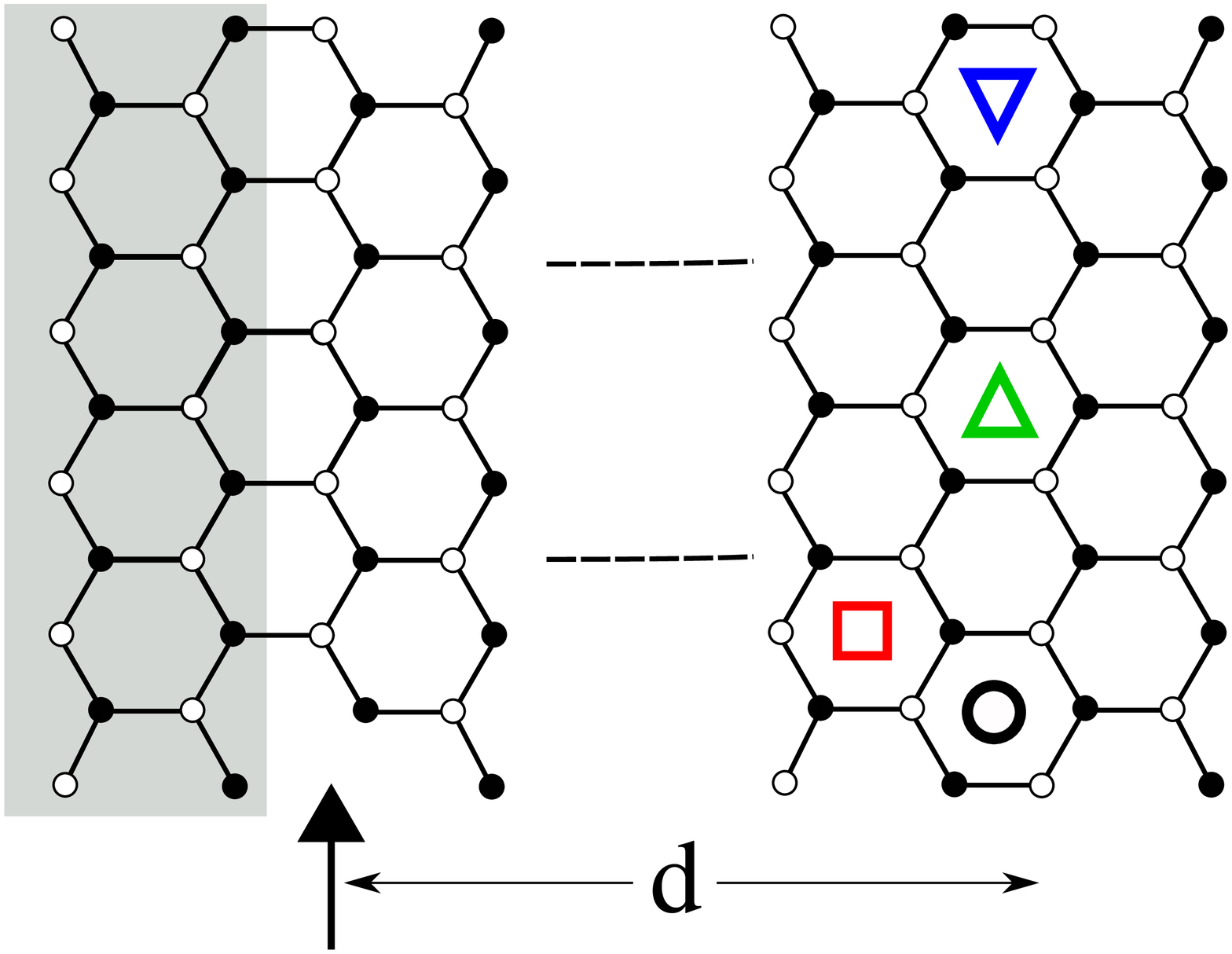} \\
\end{tabular}

\vspace{0.1cm}
\includegraphics[width=0.45\textwidth]{centads_fig}

\caption{The effect of an edge vacancy on the magnetic moment of centre-adsorbed magnetic impurity atoms on an 8-ZGNR and 11-AGNR. The notation in the schematics and graphs is the same as for Fig. \ref{edge1}. For the zigzag case, the result for an edge-adsorbed impurity (star, purple) is also shown. }
\label{edge2}
\end{figure}

Fig. \ref{edge1} shows the effect of an edge vacancy on the magnetic moments of substitutional impurities in a 6-ZGNR (left) and an 11-AGNR (right). To show the range of the effect we plot the relative change in the moment when an edge vacancy is introduced as a function of distance between the edge vacancy and the unit cell containing the magnetic impurity atom. Note that in this case we are plotting the fluctuation of each moment relative to its value at its current position in a system without edge defects, not relative to its value at the centre of the ribbon as was shown previously. This plot is shown for a number of possible sites for the magnetic impurity across the width of the ribbon, namely the edge atom on the same side as the vacancy, the site next to the edge, a site at the centre of the ribbon and a site at the opposite edge, as shown schematically in the upper panels. 

For the ZGNR, the first point to note is that the only sites that show a considerable change in their moments are the first two cases. The edge site has a slight reduction in the value of its moment, whereas the site next to the edge and belonging to the opposite sublattice to the edge has a significant increase in magnetic moment. However, the first point in this curve corresponds to a site neighbouring the edge vacancy. Excluding this, the largest deviation in magnetic moment does not exceed 5\%. However, for all positions, the moment reverts very quickly back to its value without the vacancy when it is moved further away down the ribbon. This suggests that a single edge vacancy will have very little effect on the moments of magnetic impurities located more than a lattice spacing or two away. The AGNR case is quite similar. Moving away from the vacancy the deviations in the moments again become very small. It is clear that significant deviations in the moments of substitutional impurities are not seen outside the immediate vicinity of the edge vacancy in either ribbon geometry. Similar results are shown for the adsorbed cases in Fig. \ref{edge2}. The effect here is even smaller than for the substitutional case, with fluctuations of less than 2\% at distances greater than two unit cells away from the edge vacancy for all impurity types considered, including the edge-adsorbed case in ZGNRs. 

\begin{figure}
\centering
 \hspace{0.6cm}\includegraphics[width=0.28\textwidth]{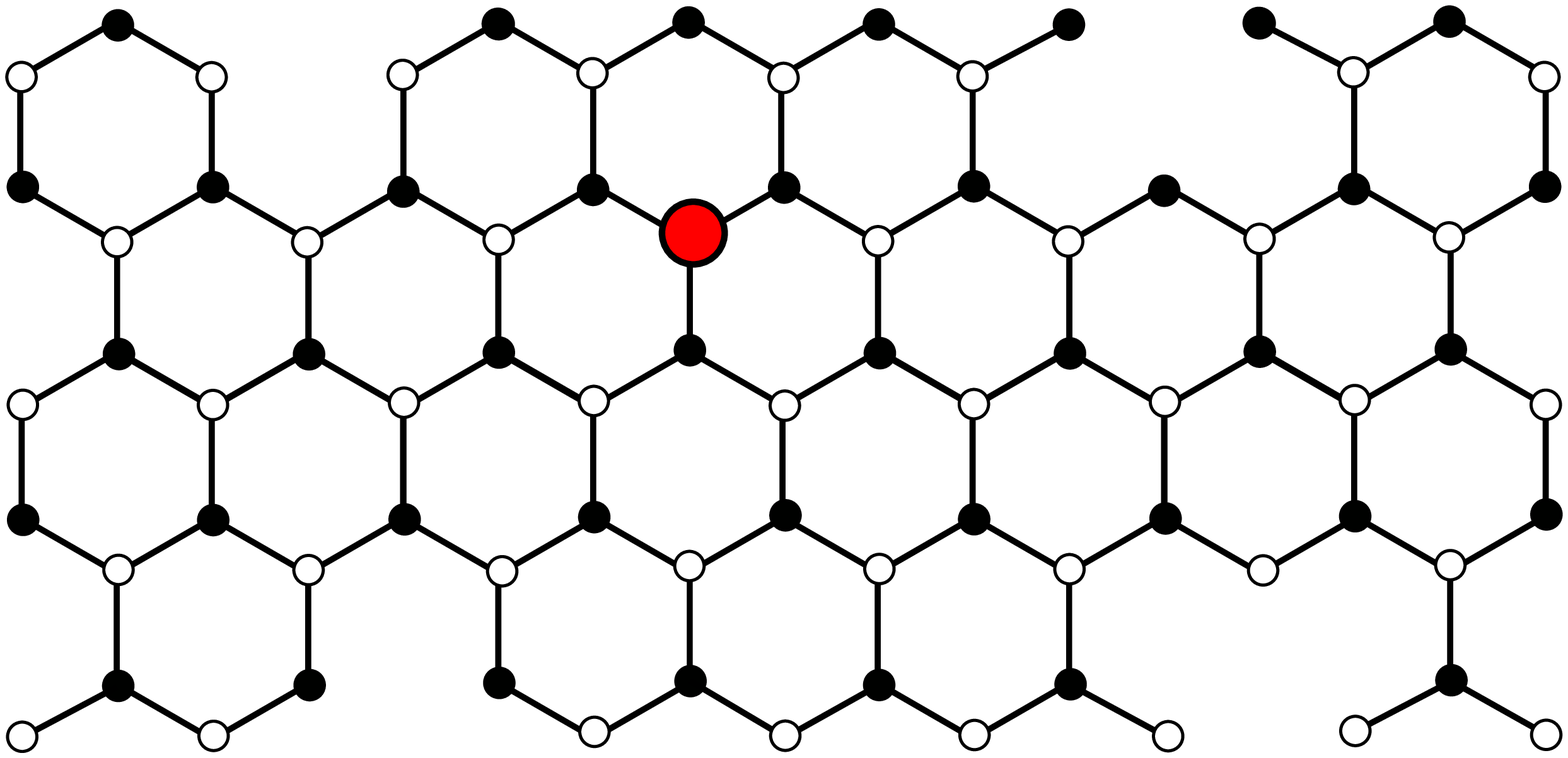}

\vspace{0.7cm}

 \includegraphics[width=0.42\textwidth]{edgedis}
\caption{Schematic showing region of 6-ZGNR with a magnetic impurity and edge disorder consisting of atoms removed randomly from the edge zigzag chain of the ribbon (top panel) and the the resulting moment fluctuations (bottom panel). The red squares indicate the position dependent moment fluctuations in a ribbon without edge disorder, whilst the black squares correspond to the average fluctuations taken over fifty edge-disorder configurations, with the standard deviation shown by the error bars. }
\label{figedgedis}
\end{figure}

A single edge vacancy has been shown not to have a significant effect on the magnetic moments of transition metal impurities in a GNRs. In fact, even the introduction of an extended edge defect, consisting of a length of ribbon to either side of the magnetic impurity with a certain concentration of edge vacancies, does not considerably affect the impurity moments unless there is an edge vacancy in their immediate vicinity. We conclude that magnetic moments introduced into GNRs by transition metal doping are particularly stable and robust in the presence of edge disorder. In particular, the striking moment profiles seen for magnetic impurities in ZGNRs will not be significantly perturbed by the introduction of a reasonably strong extended edge disorder. This point is illustrated quite clearly in Fig. \ref{figedgedis} for the case of substitutional impurities on a 6-ZGNR, the same case considered in the upper panel of Fig. \ref{figsubs}. The moment profile for the pristine case is shown as calculated with the mean-field Hubbard approach (red squares). Also shown is the moment profile for a system with a disordered region with a length of 100 unit cells to either side of the magnetic impurity (black circles). Within this disordered region, carbon atoms from the edge zigzag chain at either edge of the ribbon are removed with a probability of 10\%. This plot shows the result of an average over 50 such configurations, with the error bars on each point indicating the standard deviation. The moment profile is seen to not vary significantly from that of pristine case, demonstrating clearly the robustness of the moment profile. 

\section{Conclusions}
We have demonstrated the features of the magnetic moment variation of a magnetic impurity as its location is varied across the width of a graphene nanoribbon. For zigzag-edged nanoribbons we found an excellent agreement between the simple self-consistent Hubbard model and a more complete \emph{ab initio} treatment. Furthermore the qualitative features of the resulting moment profile remained constant for different parameterisations describing the magnetic impurity, suggesting that they hold for a wide range of magnetic species. For substitutional impurities, a nonmonotonic behaviour connected to the sublattices of the graphene atomic structure was identified. For this type of impurity, a larger moment was found on impurities located on the edge site of a ZGNR. For impurities adsorbed onto the centre of a hexagon of the graphene lattice, a monotonic increase of the moment magnitude as the impurity was moved towards the centre of the ribbon was found. However an additional impurity type, consisting of an impurity atom connecting to three edge atoms at a zigzag edge was found to have a larger moment than one connected to the edge hexagon. It was also noted to be more energetically favourable. For armchair-edged nanoribbons, the moment profile features were noted to be less robust than for the zigzag case. However the fluctuations of the moment value around that at the ribbon centre were also found to be smaller. For both edge geometries and impurity configurations, we showed that an edge vacancy did not have a significant effect on the moment of a magnetic impurity located more than one or two lattice spacings away. Furthermore, we demonstrated that the distinctive moment profile for substitutional impurities on a zigzag-edged ribbon was robust in the presence of an extended edge disorder. In light of these findings, we argue that the magnetically-doped nanoribbons may provide a route to applications previously envisaged for nanoribbons with intrinsic magnetic ordering, which is less stable in the presence of experimentally imposed constraints such as imperfect edge geometry.

\end{document}